\documentclass[global,twocolumn,hyperref]{svjour-arXiv}

\usepackage{graphics}
\usepackage{here}
\usepackage{textgreek}
\usepackage{cite}

\begin{document}
\title{Second-order Nonlinear Optical Microscopy of Spider Silk\\
}
\author{\and{Yue Zhao}\inst{1}\hbox{\href{http://orcid.org/0000-0002-8550-2020}{\includegraphics{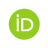}}} 
\and Khuat Thi Thu Hien\inst{1} 
\and Goro Mizutani\inst{1}\thanks{\emph{Goro Mizutani:} mizutani@jaist.ac.jp}\hbox{\href{http://orcid.org/0000-0002-4534-9359}{\includegraphics{orcid.eps}}}
\and Harvey N. Rutt\inst{2}
}                     
%
%

\institute{School of Materials Science, Japan Advanced Institute of Science and Technology, Asahidai 1-1 Nomi, 923-1292, Japan \and School of Electronic and Computer Science, University of Southampton, SO17 1BJ, UK}

\date{
\\
\\
{\it Appl. Phys. B}, {\bf 123,} 188 (2017).  DOI: 10.1007/s00340-017-6766-z}

\maketitle
\begin{abstract}
\bf Asymmetric \textbeta -sheet protein structures in spider silk should induce nonlinear optical interaction such as second harmonic generation (SHG) which is experimentally observed for a radial line and dragline spider silk by using an imaging femtosecond laser SHG microscope.  By comparing different spider silks, we found that the SHG signal correlates with the existence of the protein \textbeta -sheets.  Measurements of the polarization dependence of SHG from the dragline indicated that the \textbeta -sheet has a nonlinear response depending on the direction of the incident electric field.  We propose a model of what orientation the \textbeta -sheet takes in spider silk.
\end{abstract}
\section{Introduction}
\label{intro}
The main component of natural spider silk is the protein fibroin.  Orb-weaver spiders have seven types of secretory glands as shown in Table \ref{table1}, and the seven different types of glands produce a wide variety of gland-specific silks with different compositions and material properties.  The dragline in particular is a novel material with extreme mechanical properties having potential applications \cite{g12-39,g3-17,g3-18,g3-20,g3-16-33,g8-6,g8-93,g3-16}. 
\begin{table}[h]
\caption{Gland, function and structural component \cite{g3-16,g14,g10-1,g1-6,g8-16,g9-4,g9,g12}}
\label{table1}       
\begin{tabular}{llc}
\hline\noalign{\smallskip}
\multicolumn{1}{c}{Gland} & \multicolumn{1}{c}{Function} & Component \\
\hline\noalign{\smallskip}
Large ampullate & Dragline, Frame line & MaSp1,MaSp2 \\
Small ampullate & Radial line & MaSp1,MaSp2 \\
Flagelliform & Spiral line & Flag \\
\hline\noalign{\smallskip}
Aggregate & Sticky substance&  \\
Piriform & Attachment &  \\
Aciniform & Swathing band &  \\
Cylindrical & Egg cocoon &  \\
\hline
\end{tabular}
\end{table}

\begin{figure*}[t]
\begin{minipage}[t]{18cm}
\resizebox{1\textwidth}{!}{%
  \includegraphics{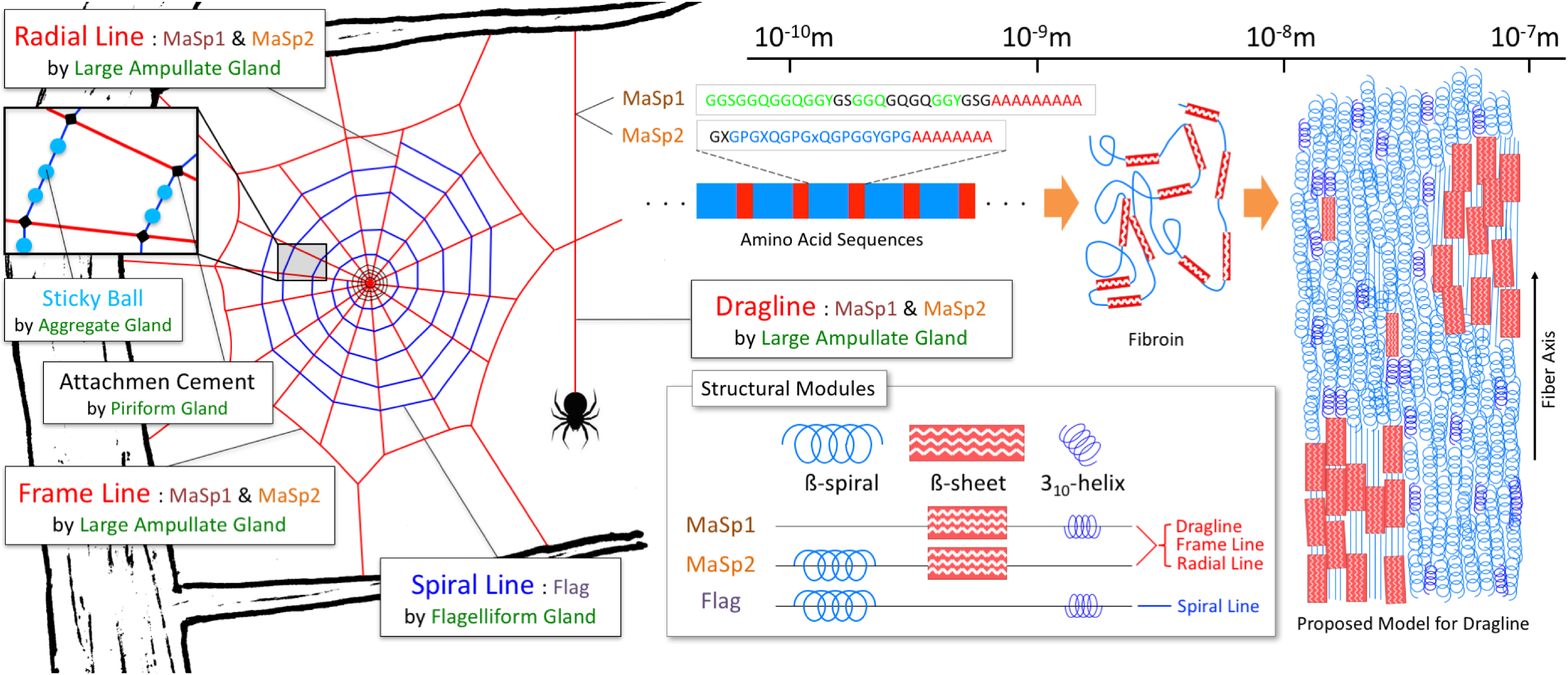}}
\end{minipage}%
\caption{Schematic of the spider's orb web, structural modules, and spider silk structure.  On the left is shown a schematic drawing of an orb web.  The red lines represent the dragline, radial line and frame lines, the blue lines the spiral line, and the center of the orb web is called the \textquotedblleft hub\textquotedblright  .  Sticky balls drawn in blue are made at equal intervals on the spiral line with viscous material secreted from the aggregate gland.  Attachment cement secreted from the piriform gland is used to connect and fix different lines \cite{g12}.  Microscopically the spider silk secondary structure is formed of fibroin and is said to have the structure  \cite{g12-39,g12,g11}  shown on the right side.  In the dragline and radial line, a crystalline \textbeta -sheet and an amorphous helical structure are interwoven \cite{g3-17,g8-39,g8-36,g8-40,g3-16-43,g2-12,g3-16-40,g2-14}.  The large amount of \textbeta -spiral structure gives elastic properties to the capture part of the orb web \cite{g3-16,g8-81}.  In the structural modules diagram \cite{g2-12} a microscopic structure of dragline and radial lines are shown, composed mainly of two proteins of MaSp1 and MaSp2 shown in the upper center part.  In the spiral line, there is no crystalline \textbeta -sheet region \cite{g2-12}.}
\label{orbweb}      
\end{figure*}

The dragline forms the structural skeleton in the spider orb web, and also serves as a lifeline for the spider. The dragline has good strength and extensibility.  For example, the strength of the spider dragline is comparable to Kevlar (poly-paraphenylene terephthalamide), and its strength is higher than steel of the same weight, whilst the energy required to cut the dragline is larger than Kevlar  \cite{g3-17,g3-16}.  The tensile strength and elongation of different kinds of line are also very different, and the elongation of spiral line is 40 times larger than that of the radial line  \cite{g3-17,g3-16}.  The fibroin molecules constituting the dragline have \textbeta -sheet structures.  In it the amino acids are arranged in a very regular order \cite{g8-39,g8-47}, and small crystal blocks \cite{g8-93,g3-30} and irregular \textquotedblleft random coils\textquotedblright appear alternately \cite{g2-15,g8}.  As shown in Fig. \ref{orbweb}, a non-crystalline region composed of a helical structure is stretched like a spring when an external force is applied, giving an elastic characteristic to the dragline \cite{g12,g12-41,nm3,nm1}, and the \textbeta -sheet structure gives high strength properties \cite{g12-39,g3-20,g12,g3-30,g3-31,g3-30-13,nm5}.  It is known that the \textbeta -sheets of the protein make the material SHG-active \cite{g16,g4}.  It was reported that SHG was detected from natural cocoon silk fibers containing a highly oriented \textbeta -sheets structure \cite{g16,g4}.  SHG was not detected from cast films made of fibroin in hexafluoro-2-propanol solutions extracted from cocoons, but was detected after forced orientation of the \textbeta -sheets structure by stretching.

As the silk is secreted, a few nanometer \textbeta -sheet crystals self-organize to form micelles of several tens of nanometers in the large ampullate gland, and then the micelles transform into a metastable liquid crystalline structures \cite{g8}.  In the spider's silk gland, the fibroin is in a high-concentration liquid crystal state in an acid environment \cite{g8-49}, and there is no birefringence \cite{g3-16-8}.  Some fibroin molecules are aligned approximately parallel to the long axis direction of neighboring molecules, due to mechanical frictional force when passing through the thin duct.  Partial ordering occurs along the fiber axis \cite{g3-16,g2-15,g8,g3-29}, and the fiber exhibits birefringence \cite{g8,g3-16-8}.

In the previous studies by X-ray diffraction (XRD) \cite{g8-92,g21,g3-16-30,g22} and nuclear magnetic resonance (NMR) \cite{g11,g8-39,g8-47,g3-16-30,g3-16-39} analysis, there were some reports that the planes of the \textbeta -sheet in the silk were oriented almost parallel to the fiber axis.  However, it is still not perfectly known what orientation the \textbeta -sheet takes in spider silk.  The XRD and NMR measurement has very poor spatial resolution, and information on the microstructure distribution in the sample was not obtained.

On the other hand, since the second-order nonlinear optical effect is sensitive to the location and orientation of asymmetric structures, SHG microscopy can be used to detect them efficiently with good spatial resolution.  However, there has been no report on the second-order nonlinear optical properties of spider silk to our knowledge.  In this study, the second-order nonlinear optical effect of spider silk was directly observed rather than that of cocoon fibroins by using a polarization-resolved femtosecond laser SHG microscope for the first time to our knowledge.

\section{Method and Sample}
\label{sec:Method and Sample}
\subsection{Collection of Samples}
\label{sec:Collection of Samples}
\begin{figure}[h]
\begin{minipage}[t]{8.5cm}
\resizebox{1\textwidth}{!}{  \includegraphics{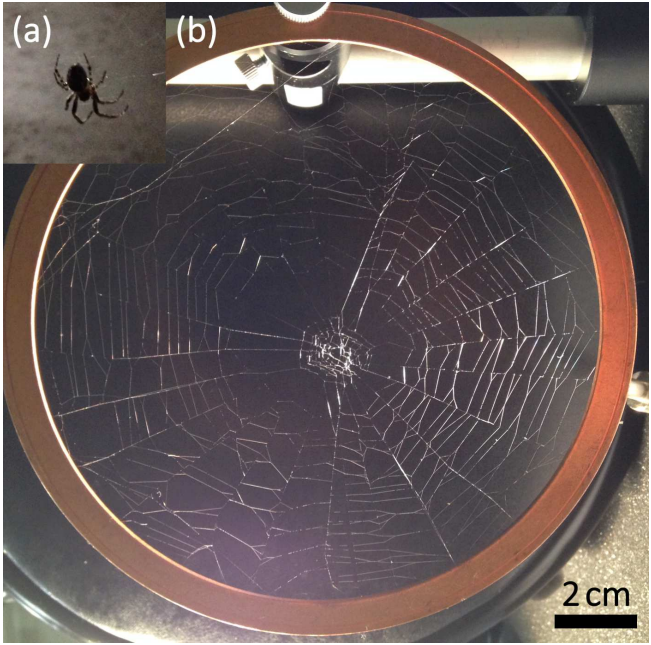}}
\end{minipage}%
\caption{(a) A spider (\textit{Araneus ventricosus}).  (b) The sample of an orb web and the holder.}
\label{sample}  
\end{figure}

The samples are an orb web and a dragline.  The latter was wound up from a living spider (\textit{Araneus ventricosus}) (see Fig. \ref{sample}(a)).  In order to collect draglines, we put the spider on an aluminum frame, and gave it a small jerk, causing it to fall lightly and produce a dragline. We immediately wound the dragline and collected it while rotating the frame so that the dragline was not cut and the spider did not fall to the ground.  In order to take the orb web of a wide area, we let the spider leave its orb web, and then we picked up the orb web with a large copper ring (see Fig. \ref{sample}(b)).

\subsection{Observation of Samples}
\label{sec:Observation of Samples}

\begin{figure}[h]
\begin{minipage}[t]{8.5cm}
\resizebox{1\textwidth}{!}{  \includegraphics{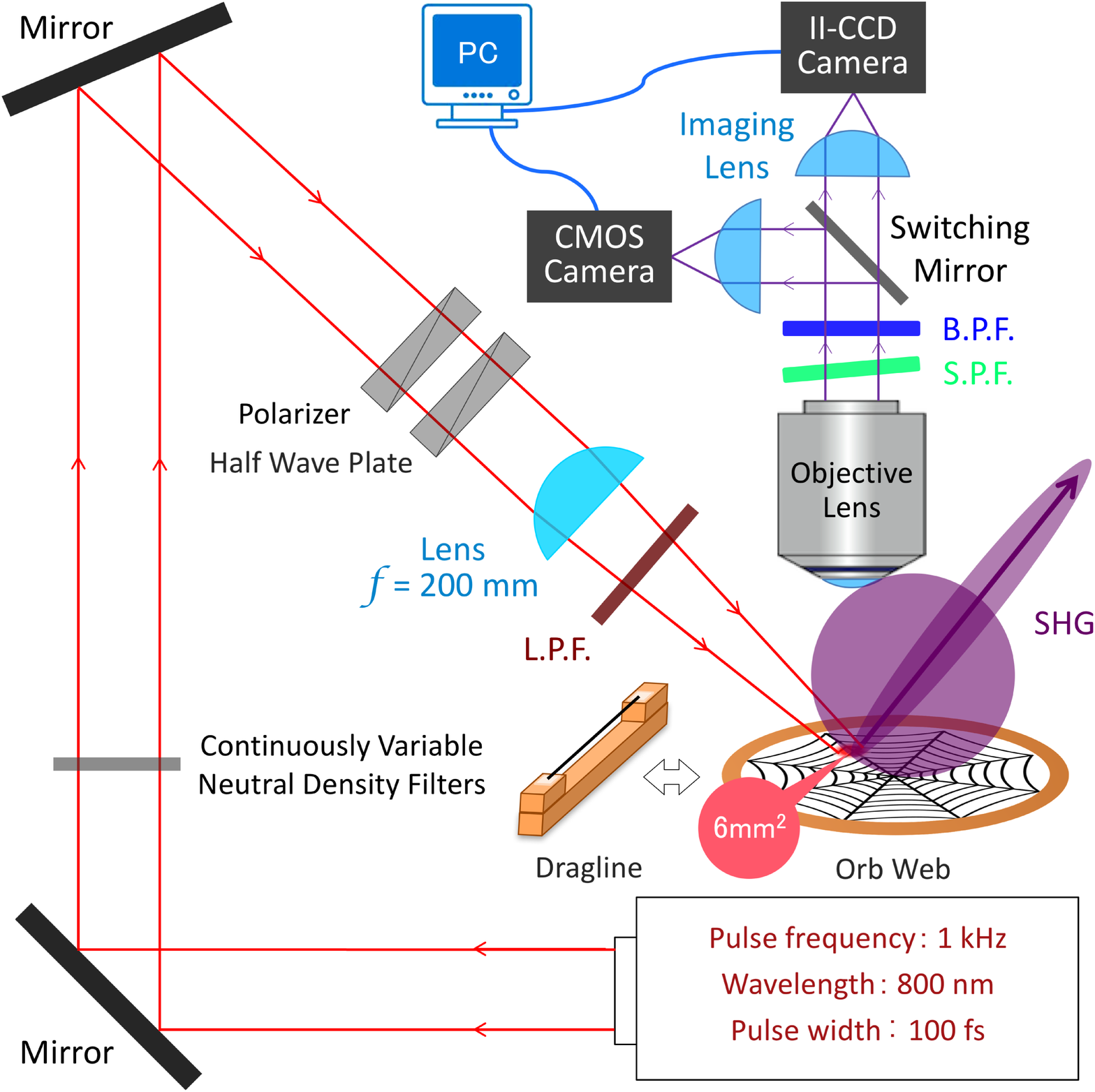}}
\end{minipage}%
\caption{Experimental setup.  SHG microscopy system.  A CMOS camera (Lumenera Corp., Lu135M) was used for aiming.  The light source was a femtosecond pulse laser.  The laser beam was focused loosely on the sample, and the incident angle with respect to the normal to the sample stage was 60$^\circ$.  Scattered SHG light from the sample was imaged by an objective lens and detected by an image intensified charge coupled device (II-CCD) camera (HAMAMATSU, PMA-100).  When observing 2PEF images, we only replaced the bandpass filter of 400 nm with one of 438 nm.  The short wavelength pass filter is tilted by 5$^\circ$ with respect to parallel rays to eliminate \textquoteleft ghost\textquoteright \  signals.  L.P.F.: Longwavelength pass filter.  S.P.F.: Shortwavelength pass filter.  B.P.F.: Bandpass filter.
}
\label{setup}  
\end{figure}

The SHG of the sample was observed in a scattering geometry \cite{method}.  The setup of the optical system to observe the SHG image of the sample is shown in the Fig. \ref{setup}.  When observing the orb web, we put the copper ring on the stage directly.  In the case of observing a dragline, both ends of the dragline were fixed with double-sided tape to a wooden holder.  The light source was a pulse train with a repetition frequency of 1 kHz, wavelength of 800 nm, and a pulse width of about 100 fs.  A seed light source (mode-locked Ti:sapphire laser, Spectra-Physics: Tsunami) was amplified by a regenerative amplifier (Spectra-Physics: Spitfire).  The excitation power at the sample position was controlled by a continuous variable neutral density filter.  The incident angle with respect to the normal to the sample stage was 60$^\circ$.  In order to make the irradiated area about 6 mm$^2$, the sample was placed at 10 mm off the focal point of the focusing lens (focal length $f$ = 200 mm).  Since the irradiated area is much larger than the view field of the camera, the distribution of the irradiating light in the observed field is substantially uniform.  Before the condenser lens, we put a linear polarizer.  After the polarizer, we put a half wave plate to rotate the polarization on the sample.  Between the focusing lens and the sample, long wavelength pass filter (L.P.F.) was used to remove light of wavelength shorter than 780 nm, and pass the 800 nm beam.  

The imaging optics was a commercial microscope OLYMPUS BX60.  First, we used a CMOS camera (Lumenera Corp., Lu135M) for alignment.  Then, the sample was illuminated by the laser pulses.  SHG light from the sample passed through the objective lens, became a parallel ray (infinity-corrected optics), passed through a \textquoteleft Semrock\textquoteright \  short wavelength pass filter BSP01-785R, rejecting 800 nm wavelength, and finally was selected by a \textquoteleft Semrock\textquoteright \  bandpass filter FF01-395/11 with a center wavelength of 400 nm.  The transmittance curve of FF01-395/11 is rectangular shaped and drops sharply to zero at 387.8 nm and 402.6 nm.  When checking the image of the 2PEF, we replaced the bandpass filter with an alternative \textquoteleft Semrock\textquoteright \  band pass filter FF02-438/24 with a center wavelength of 438 nm.  The SHG and 2PEF images were observed using the photon counting function of an image intensified - charge coupled device (II-CCD) camera (HAMAMATSU, PMA-100).  The spatial resolution of the microscope is decided by the chip size of the II-CCD camera, 11 \textmu m $\times$ 13 \textmu m, and was 2.6 \textmu m for magnification $\times$5 (NA=0.15) and 0.65 \textmu m for magnification $\times$20 (NA=0.46).  

In this study, the excitation light energy density of one pulse was 13 \textmu J/mm$^2$. In order to check the damage threshold, the excitation light energy density of one pulse was raised to 33 \textmu J/mm$^2$ with the same optical setup, and the irradiation was maintained for an hour, but no damage was observed visually, and the SHG image did not change after this test.  Therefore, damage to the sample is considered to be negligible.  Each measurement required an irradiation time of 6 min at maximum.

\section{Result and Discussion}
\label{sec:Result and Discussion}
\subsection{SHG of Orb Web}
\label{sec:SHG of Orb Web}

\begin{figure*}[ht]
\begin{minipage}[h]{18cm}
\resizebox{1\textwidth}{!}{%
  \includegraphics{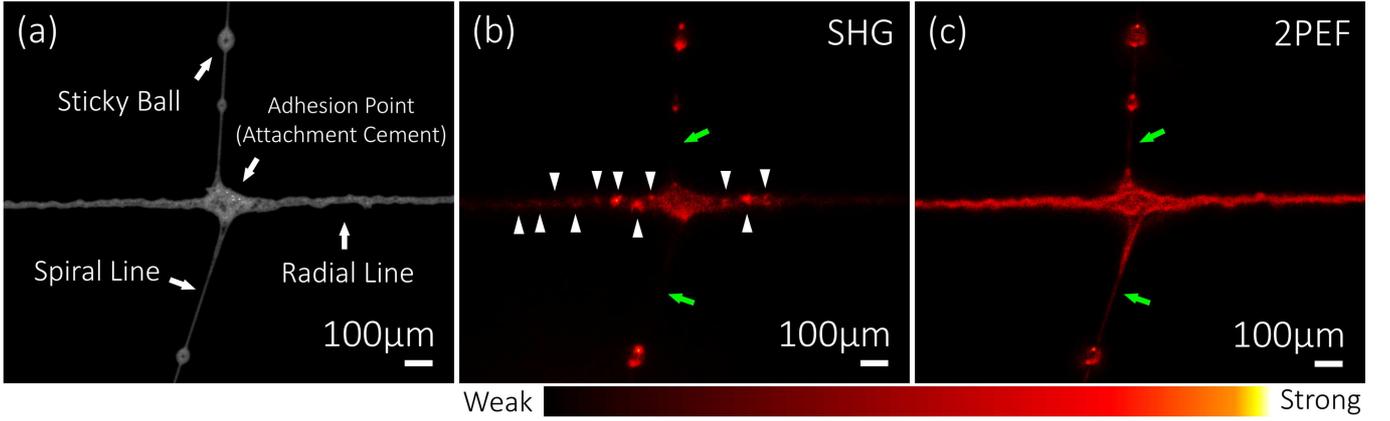}}
\end{minipage}%
\caption{(a) Microscopic image of a spider orb web taken by a CMOS camera with white light illumination.  The radial line, spiral line (sticky balls on spiral line) and the adhesion point of two lines can be seen.  (b) SHG image and (c) 2PEF image.  (b) and (c) were observed with the bandpass filter at 400 nm and 438 nm wavelengths, respectively.  The magnification of the objective lens was $\times$5 (NA = 0.15).  For (b) and (c), the incident light wavelength was 800 nm, and the energy density was 13 \textmu J/mm$^2$. The integration time of (b) was 300 s, and that of (c) was 50 s.}
\label{orbweb_shg}      
\end{figure*}

Figure \ref{orbweb_shg}(a) is a linear optical microscopic image of a part of the orb web of \textit{Araneus ventricosus} with white light illumination.  Figure \ref{orbweb_shg}(b) is an intensity image observed with an II-CCD camera attached to the microscope with a for 400 nm centre bandpass filter corresponding to SHG behind the objective lens when the laser pulses illuminated the sample.  Figure \ref{orbweb_shg}(c) is an intensity image observed by replacing the bandpass filter with the one at 438 nm corresponding to 2PEF.  If the signals in Fig. \ref{orbweb_shg}(b) and Fig. \ref{orbweb_shg}(c) are due to the same origin (broad spectral width, e.g., luminescence or background), similar images should be obtained at both wavelengths of 400 nm and 438 nm.  However, in Fig. \ref{orbweb_shg}(b) and Fig. \ref{orbweb_shg}(c), the positions of the intense signals are different.  Namely, in Fig. \ref{orbweb_shg}(b), some enhanced SHG signals of several tens of micrometers at periodic positions (white marks in Fig. \ref{orbweb_shg}(b)) are observed.  On the other hand, the spiral line secreted from the flagelliform gland showed 2PEF signals, but almost no SHG signal (Fig. \ref{orbweb_shg}(c) and (b), green marks).  This is the first observation of SHG from the radial line of the spider orb web to our knowledge.  A strong 2PEF signal from the sticky balls attached to the spiral line is detected in both Fig. \ref{orbweb_shg}(b) and Fig. \ref{orbweb_shg}(c), and it is considered to be luminescence.

We tried to evaluate the statistically averaged value of $\chi^{(2)}$ elements of the spider silk of Fig. \ref{orbweb_shg}(b) using a reference zinc sulfide polycrystalline pellet by powder technique of Kurtz et al.\cite{powder}  We chose the radial line because its refractive index is known\cite{SS_RI} and the coherence length can be evaluated.  We evaluated the $\chi^{(2)}$ of the SHG spots in Fig. \ref{orbweb_shg}(b).  According to the powder technique of Kurtz et al.\cite{powder} the intensity of SHG is given by the following expression.
\begin{eqnarray}
\label{eq1}
{ I }^{ 2\omega  }\propto \frac { L\hat { r }  }{ { { \hat { l }  }_{ c } } ^{ 2 }} \left< { \left( { d }^{ 2\omega  } \right)  }^{ 2 } \right>
\end{eqnarray}
Here ${I}^{2\omega}$ is the intensity of SHG, $L$ is the absorption length, $\hat{r}$ is the average size of powder particle, $\hat{l}_{c}$ is the coherence length, and $\left<({d}^{2\omega})^{2}\right>$ is the average by the molecular orientation of the square of the second order nonlinear susceptibility ${d}^{2\omega}$.  The coherence length in SHG can be calculated from the refractive index $n_{2\omega}$ at the SHG light wavelength, the refractive index $n_{\omega}$ at the fundamental light wavelength, and the wavelength $\lambda$ of the fundamental light as,
\begin{eqnarray}
\label{eq2}
{ \hat { l }  }_{ c }\equiv { \left< \frac { \lambda  }{ 4\left( { n }_{ 2\omega  }-{ n }_{ \omega  } \right)  }  \right>  }_{ av } .
\end{eqnarray}
First, the powder particle size $\hat{r}$ and the absorption length $L$ can be regarded as the spider silk line diameter of 17.8 \textmu m.  The refractive index of radial line is $n_{2\omega}$=1.560-1.583 at wavelength 400 nm and $n_{\omega}$=1.536-1.575 at wavelength 800 nm \cite{SS_RI}.  Here, the refractive indices are the values of the radial lines from {\it Plebs eburnus}'s orb web \cite{SS_RI}.  Substituting these into equation (\ref{eq2}), the coherence length of the radial line is calculated as $\hat{l}_{c}$= 8-25 \textmu m.  Since the coherence length and the diameter of the radial line have nearly the same values, the SHG light was almost phase matched.  The average domain size of the ZnS crystals was $\hat{r}$=1 \textmu m \cite{ZnS_size}.   The absorption length $L$ at 400 nm wavelength is 596-1216 \textmu m as calculated from the known optical density \cite{ZnS_OD}.  The refractive indices of ZnS are $n_{2\omega}$=2.57 at wavelength 400 nm and $n_{\omega}$=2.31 at wavelength 800 nm \cite{ZnS_RI}.  Substituting these values into equation (\ref{eq2}), the coherence length of ZnS is calculated as $\hat{l}_{c}$=0.79 \textmu m.  The second-order nonlinear susceptibility $\left|{d}^{\left(2\right)}\right|$ of zinc sulfide is 18.8 pm/V at the fundamental wave of 800 nm \cite{ZnS_Chi2}.  Using the relation ${ \chi  }^{ \left( 2 \right)  }=2{ d }^{ \left( 2 \right)  }$ \cite{Shen_PNO}, the value $\sqrt{\left<|{\chi}^{(2)}|^{2}\right>}$ of the second-order nonlinear susceptibility of zinc sulfide at the fundamental wave of 800 nm is 37.6 pm/V.  At the fundamental wave of 800 nm our result was $I^{2\omega}_{\rm Spider Silk}$/$I^{2\omega}_{\rm ZnS}$=1.3$\times$10$^{-6}$.  According to equation (\ref{eq1}), the  ${ \chi  }^{ \left( 2 \right)  }=2{ d }^{ \left( 2 \right)  }$ \cite{Shen_PNO}, the value $\sqrt{\left<|{\chi}^{(2)}|^{2}\right>}$ of the radial line is calculated as 0.6-2.7 pm/V.

\subsection{SHG of Drag Line and Incident Polarization Dependence}
\label{sec:SHG of Drag Line and Its Incident Polarization Dependence}

\begin{figure*}[ht]
  \begin{minipage}[ht]{18cm}
\resizebox{1\textwidth}{!}{  \includegraphics{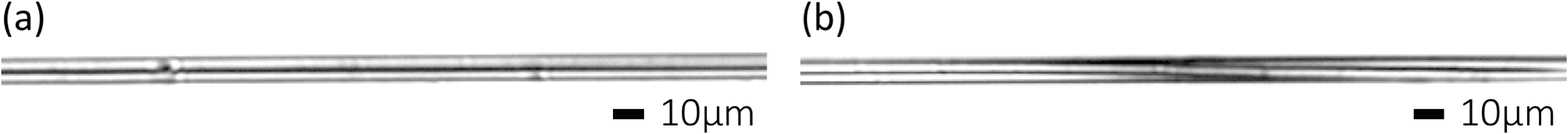}}
\hspace{-5mm}
\caption{Transmission light bright field optical microscopic images of pairs of draglines with white light illumination taken by a CMOS camera.  (a) Two parallel draglines.  (b) Two twisted draglines.}
\label{dragline_cmos} 
\end{minipage}%

\bigskip

\begin{minipage}[ht]{18cm}
\resizebox{1\textwidth}{!}{%
  \includegraphics{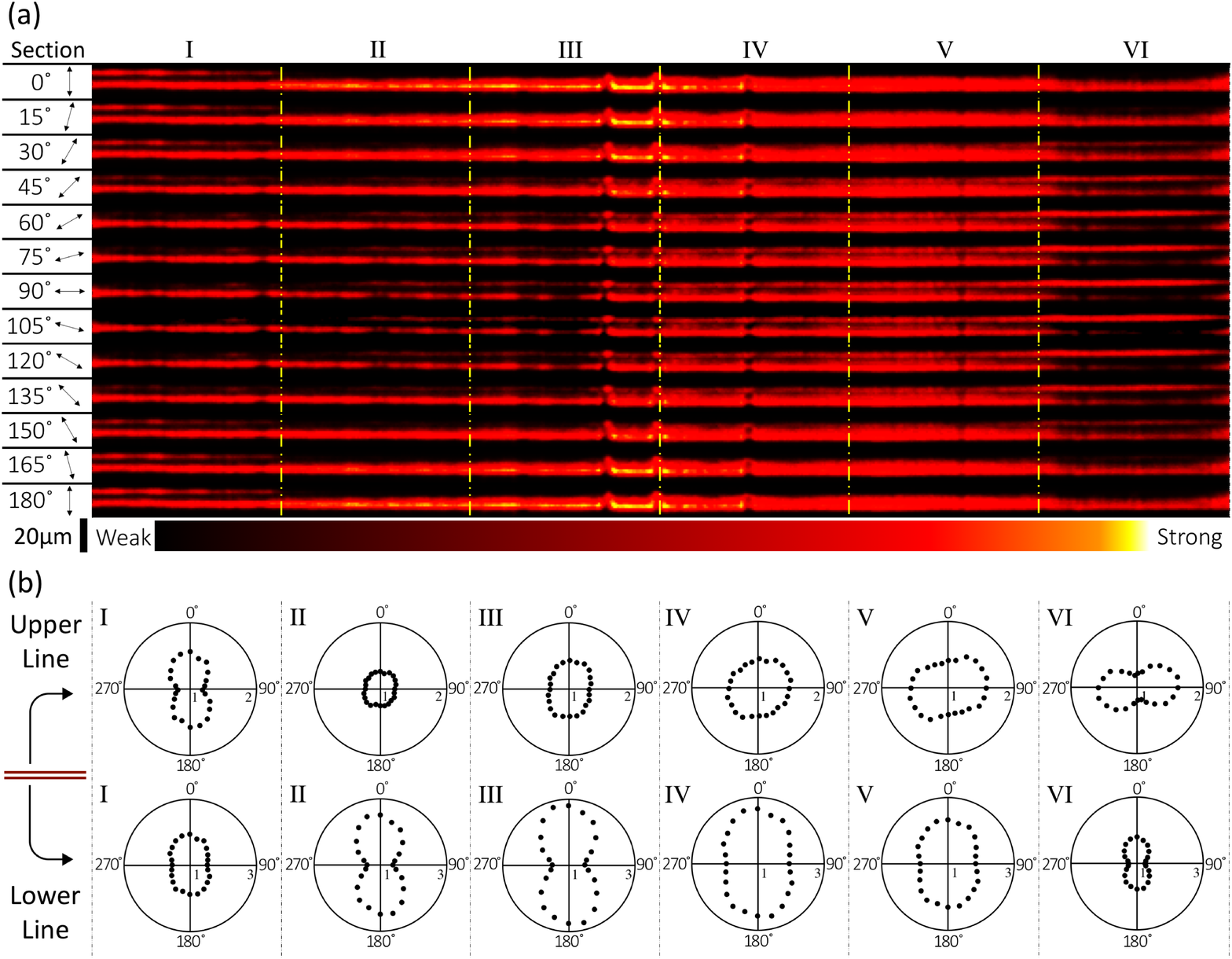}} 
\caption{(a) Dependence of the SHG image on the incident light polarization.  The energy of excitation light density was 13 \textmu J/mm$^2$. The integration time was 1 s.  The magnification of the objective lens was $\times$20 (NA = 0.46).  The polarization of the incident light was rotated by each 15$^\circ$ in a clockwise direction with a half wave plate.  The polarization of observed SHG was not specified.  The angle is defined as 0$^\circ$ when the incident light electric field is directed perpendicular to the dragline fiber axis.  The incident polarization angle and icons are shown on the left side.  (b) Polar graphs of the incident light polarization dependence of SHG intensities for sections I to VI.  The top polar graphs are for the upper line and the bottom polar graphs are for the lower line. }
\label{dragline_shg_polarization}      
\end{minipage}%
\end{figure*}

The dragline is secreted from a pair of spinnerets as two lines as seen in the microscopic images in Fig. \ref{dragline_cmos}(a) and Fig. \ref{dragline_cmos}(b).  SHG images and the incident light polarization dependence of this SHG signal are shown in Fig. \ref{dragline_shg_polarization}(a) and Fig. \ref{dragline_shg_polarization}(b), respectively.  Here we chose the two parallel draglines as shown in Fig. \ref{dragline_cmos}(a) for observation. The polarization of incident light was rotated by 15$^\circ$ step, and the observed light polarization was not specified.  Unfortunately the SHG signal was very weak when we inserted a polarizer after the objective lens.  SHG was detected from the dragline also for the first time to our knowledge.  The 2PEF signal from the dragline was below the noise level (not shown), so all the signals observed in Fig. \ref{dragline_shg_polarization}(a) are SHG.

In Fig. \ref{dragline_shg_polarization}(a), the SHG intensity depends strongly on position.  In addition, the relative SHG intensity ratio of the two draglines varies as a function of the position.  The intensity of the SHG at one position varies as a function of the incident light polarization angle.  Figure \ref{dragline_shg_polarization}(b) shows the polar graphs of SHG intensities as a function of the incident light polarization angle sampled from section I to VI of the upper and lower lines.  In Fig. \ref{dragline_shg_polarization}(b), the intensities from 0$^\circ$-180$^\circ$ are the measured values, and those from 180$^\circ$-360$^\circ$ are copies of 0$^\circ$-180$^\circ$ values, since the polarizations at 180$^\circ$-360$^\circ$ are equivalent to 0$^\circ$-180$^\circ$.  The two-lobed pattern in the polar graphs of the upper line on the top row at section I gradually becomes circular to section IV, and finally at section VI, becomes one rotated by 90$^\circ$ from that of section I.  The pattern in the polar graphs of the lower line on the bottom row expands in the vertical direction from sections I to III, and shrinks again from section IV to VI.  But, the central constrictions of the pattern became less distinct in sections IV and V.  In sections III and IV in Fig. \ref{dragline_shg_polarization}(a), SHG spots of several micrometers size were observed on the upper line.  The intensities of these SHG spots also depend on the incident light polarization.

\subsection{Discussion}
\label{sec:Discussion}

In Fig. \ref{orbweb_shg}(b) spot-like SHG images were observed on the radial line, but not on the spiral line.  In Fig. \ref{dragline_shg_polarization}(a) SHG was observed from the entire dragline.  The only difference between the constituents among the dragline, radial line, and the spiral line is the \textbeta -sheet of the protein.  Namely, only the spiral line contains no \textbeta -sheets \cite{g2-12}.  As mentioned in the introduction, SHG is generated from natural cocoon silk fibers containing a highly oriented \textbeta -sheets structure or from a film with \textbeta -sheets structure oriented by stretching \cite{g16}.  Therefore, it can be inferred that the observed SHG of the radial line and the dragline originates from oriented protein \textbeta -sheets.

The \textbeta -sheets in silk fibroin of cocoons are oriented by stretching \cite{g19}.  As shown in the \textquotedblleft structural modules\textquotedblright \  box in Fig. \ref{orbweb}, the dragline as well as the radial line consists of proteins MaPs1 and MaPs2 and they contain modules called \textbeta -sheets.  The \textbeta -sheet is oriented as it passes through a narrow duct of the spider's gland, and is thought to thereby constitute an anisotropic crystal region having birefringence \cite{g3-16,g8,g3-16-8}.  There was a report that the \textbeta -sheets in a dragline were oriented almost parallel to the fiber axis \cite{g3-16}.  Therefore, oriented \textbeta -sheet nanocrystals in the radial line and dragline are considered to have induced SHG as a macroscopic non-centrosymmetric structure.  When we took the dragline from a spider, the dragline was pulled by the weight of the spider.  When the orb web was made, the spider pulled the silk to stretch it.  Therefore, taking the result of Ref. \cite{g16} into account, it is considered that \textbeta -sheets were oriented by the external force. 

\begin{figure}[b]
\begin{minipage}[t]{8.5cm}
\resizebox{1\textwidth}{!}{  \includegraphics{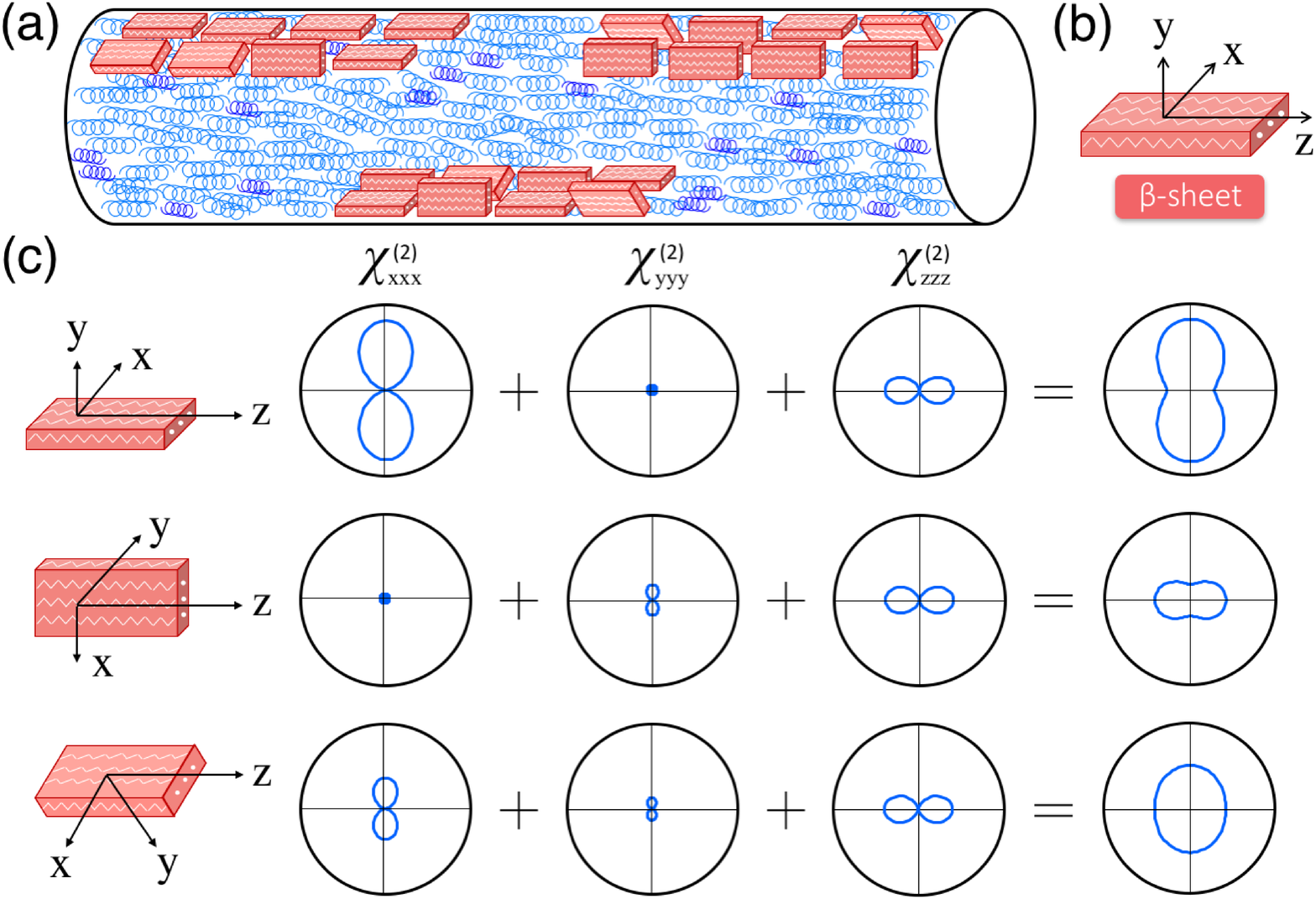}}
\end{minipage}%
\caption{(a) Schematic representation of proposed model for \textbeta -sheets orientation of a spider-silk dragline.  (b) Definition of x, y, z direction on the sample coordinate system of the \textbeta -sheet.  (c) SHG polar graphs of three components of nonlinear susceptibility in \textbeta -sheet and their sum.
}
\label{orientation_anisotropy}  
\end{figure}

According to past reports \cite{g3-16,g11,g8-36}, the \textbeta -sheet is inferred to be oriented in the spider's dragline as shown on the right side of Fig. \ref{orbweb}.  There is no information on the anisotropy of nonlinear susceptibility ${\chi}^{(2)}$ of the \textbeta -sheet in the spider's silk.  So, considering the results of the incident light polarization dependence in this study, we propose a tentative model as follows.  Figure \ref{orientation_anisotropy}(b) defines x, y, z-direction of the crystallographic coordinate system of the \textbeta -sheet.  Here as one tentative explanation we assume three nonzero components of the nonlinear susceptibility ${\chi}_{\rm xxx}^{(2)}$, ${\chi}_{\rm yyy}^{(2)}$,  ${\chi}_{\rm zzz}^{(2)}$ (${\chi}_{\rm xxx}^{(2)} \neq {\chi}_{\rm yyy}^{(2)}$) of the \textbeta -sheets.  Namely, our experimental results cannot be explained by a simple hyperpolarizability ${\beta}_{ \xi \xi \xi }$ of a rod-like molecule with $\xi$ representing the coordinate in its molecular axis direction.  Thus we need more components of ${\chi}^{(2)}$ and the one suggested above is one of the simplest options.  In the model in Fig. \ref{orientation_anisotropy}(a), the z-axes of the \textbeta -sheets are oriented almost parallel to the fiber axis, while the x- or y-axes of the \textbeta -sheets are distributed randomly in the plane perpendicular to the fiber axis.  As a result, total ${\chi}_{\rm zzz}^{(2)}$ is constant regardless of the orientation of the \textbeta -sheet units.  On the other hand, if the values of ${\chi}_{\rm xxx}^{(2)}$ and ${\chi}_{\rm yyy}^{(2)}$ are different from each other, the total ${\chi}^{(2)}$ in the direction perpendicular to the fiber axis changes with rotation of the \textbeta -sheets.  The polarization dependence pattern should change due to the cooperative effects of ${\chi}_{\rm xxx}^{(2)}$, ${\chi}_{\rm yyy}^{(2)}$ and ${\chi}_{\rm zzz}^{(2)}$ as shown in Fig. \ref{orientation_anisotropy}(c).  This may correspond to the pattern shape variation in Fig. \ref{dragline_shg_polarization}(b).  The origin should be assigned to the variation of the density of the \textbeta -sheets, or the orientation of the \textbeta -sheets, or both.

\begin{figure}[h]
\begin{minipage}[t]{8.5cm}
\resizebox{1\textwidth}{!}{  \includegraphics{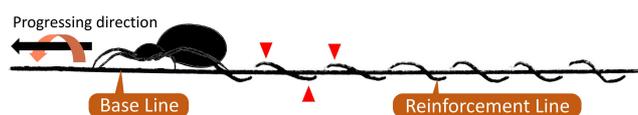}}
\end{minipage}%
\caption{Schematic illustration of the situation when a spider reinforce the radial line.  The SHG spots may have occurred at positions with red marks on the radial line.
}
\label{reinforcement_work}  
\end{figure}

The periodic SHG spots in Fig. \ref{orbweb_shg}(b) may be interpreted as follows.  We observed spiders when making their orb webs.  When the spider starts to make an orb web, it stretched radial lines as the \textquotedblleft base lines\textquotedblright \  first.  Then, it secreted silk to reinforce the \textquotedblleft base line\textquotedblright \  while advancing its body spirally around it, as illustrated in Fig. \ref{reinforcement_work}.  The first \textquotedblleft base line\textquotedblright \  takes tension while the second \textquotedblleft reinforcement line\textquotedblright \  is spirally wrapped around \textquotedblleft base line\textquotedblright  .  Here it can be inferred that SHG is most efficiently generated when the \textquotedblleft reinforcement line\textquotedblright \  is directed at a certain angle. Therefore, the periodic enhanced SHG signal in Fig. \ref{orbweb_shg}(b) as shown by the white marks is observed.

Finally, a strong signal from the viscous sphere was observed in both Fig. \ref{orbweb_shg}(b) and Fig. \ref{orbweb_shg}(c). The origin of this signal is not SHG but is due to the fluorescent property of the sticky ball itself.

\section{Summary}
\label{sec:Summary}
In summary, we have observed, for the first time to our knowledge, the SHG signal and SHG image from the spider silk of the radial line and dragline.  The second-order nonlinear susceptibility of radial line $\sqrt{\left<|{\chi}^{(2)}|^{2}\right>}$= 0.6 - 2.7 pm/V was tentatively obtained using a polycrystalline zinc sulfide pellet as a reference using the powder technique by Kurtz et al.  On the other hand, the SHG from the spiral line was below the noise level.  It can be judged that the origin of SHG is the \textbeta -sheet of protein.  The incident light polarization dependence may be caused by the arrangement of \textbeta -sheets of protein in the spider silk.

\bibliographystyle{0.zidingyi}
\bibliography{references}
\end{document}